# DYNAMICS OF NONHARMONIC INTERNAL GRAVITY WAVE PACKETS IN STRATIFIED MEDIUM

V.V.Bulatov
Institute for Problems in Mechanics
Russian Academy of Sciences
Pr.Vernadskogo 101-1, 117526 Moscow, Russia
bulatov@index-xx.ru

*Abstract*

*In the paper taking the assumption of the slowness of the change of the parameters of the vertically stratified medium in the horizontal direction and in time, the evolution of the non-harmonic wave packages of the internal gravity waves has been analyzed. The concrete form of the wave packages can be expressed through some model functions and is defined by the local behavior of the dispersive curves of the separate modes near to the corresponding special points. The solution of this problem is possible with the help of the modified variant of the special-time ray method offered by the authors (the method of geometrical optics), the basic difference of which consists that the asymptotic representation of the solution may be found in the form the series of the non-integer degrees of some small parameter. At that the exponent depends on the concrete form of representation of this package. The obvious kind of the representation is determined from the principle of the localness and the asymptotic behavior of the solution in the stationary and the horizontally-homogeneous case. The phases of the wave packages are determined from the corresponding equations of the eikonal, which can be solved numerically on the characteristics (rays). Amplitudes of the wave packages are determined from the laws of concervation of the some invariants along the characteristics (rays).*

The present paper presents the basic provisions of the method of the geometric optics or Wentzel-Kramers-Brillouin (WKB) approximations with consideration of the specificity of the internal gravity waves. If to consider the internal waves for the case, when the undisturbed field of density $\rho_0(z, x, y)$ depends not only on the depth $z$, but also on the horizontal coordinates $x$ and $y$, then, generally speaking, if the undisturbed density is the function of the horizontal coordinates, then such a density distribution forms some field of the horizontal flows. However these flows are rather slow and as a first approximation they can be neglected, and so it is usually considered, that the field $\rho_0(z, x, y)$ is set a priori and thereby one may suppose the presence of some exterior sources, or the non-conservatism of the system under analysis. It is also quite evident, that, if the internal waves are propagated above the irregular bottom, then such problem does not originate, because the system "the internal wave – the irregular bottom" is conservative and there is no inflow of the energy from the outside.

Let's further consider the linearized system of the equations of hydrodynamics:

$$\rho_0 \frac{\partial \tilde{U}_1}{\partial t} = -\frac{\partial p}{\partial x}$$
$$\rho_0 \frac{\partial \tilde{U}_2}{\partial t} = -\frac{\partial p}{\partial y}$$
$$\rho_0 \frac{\partial \tilde{W}}{\partial t} = -\frac{\partial p}{\partial z} + g\rho \qquad (1)$$
$$\frac{\partial \tilde{U}_1}{\partial x} + \frac{\partial \tilde{U}_2}{\partial y} + \frac{\partial \tilde{W}}{\partial z} = 0$$
$$\frac{\partial \rho}{\partial t} + \tilde{U}_1 \frac{\partial \rho_0}{\partial x} + \tilde{U}_2 \frac{\partial \rho_0}{\partial y} + \tilde{W} \frac{\partial \rho_0}{\partial z} = 0$$

Here $(\tilde{U}_1, \tilde{U}_2, \tilde{W})$ are the vectors of velocity of the internal gravity waves, $p$ and $\rho$ are disturbances of the pressure and density, $g$ - an acceleration of gravity ($z$-axis is directed downwards).

Using Boissinesq approximation, which means, that the density $\rho_0(z, x, y)$ in the first three equations of the system (1.1) is considered as a constant value, with the help of the crossover differentiation we shall re-arrange the system (1.1) into

$$\frac{\partial^4 \tilde{W}}{\partial z^2 \partial t^2} + \Delta \frac{\partial^2 \tilde{W}}{\partial t^2} + \frac{g}{\rho_0} \Delta(\tilde{U}_1 \frac{\partial \rho_0}{\partial x} + \tilde{U}_2 \frac{\partial \rho_0}{\partial y} + \tilde{W} \frac{\partial \rho_0}{\partial z}) = 0$$

$$\frac{\partial}{\partial t}(\Delta \tilde{U}_1 + \frac{\partial^2 \tilde{W}}{\partial z \partial x}) = 0 \tag{2}$$

$$\frac{\partial}{\partial t}(\Delta \tilde{U}_2 + \frac{\partial^2 \tilde{W}}{\partial z \partial y}) = 0$$

where $\Delta = \partial^2/\partial x^2 + \partial^2/\partial y^2$.

As the boundary conditions we shall use the condition of "the solid cover"

$$W = 0 \quad \text{at} \quad z = 0, H. \tag{3}$$

Let's consider the harmonic waves $(\tilde{U}_1, \tilde{U}_2, \tilde{W}) = e^{i\omega t}(U_1, U_2, W)$ and introduce the dimensionless variables in the formulas

$$x^* = \frac{x}{L}, \quad y^* = \frac{y}{L}, \quad z^* = \frac{z}{h},$$

where $L$ is the characteristic scale of variation of $\rho_0$ in the horizontal direction, $h$ is the characteristic scale of variation of $\rho_0$ in the vertical direction (for example, the width of the thermocline).

In the dimensionless coordinates (2) we shall have the following view (the character $*$ further down is omitted):

$$-\omega^2 (\frac{\partial^2 W}{\partial z^2} + \varepsilon^2 \Delta W) + \varepsilon^2 \frac{g_1}{\rho_0}(\varepsilon U_1 \frac{\partial \rho_0}{\partial x} + \varepsilon U_2 \frac{\partial \rho_0}{\partial y} + W \frac{\partial \rho_0}{\partial z}) = 0,$$

$$\varepsilon \Delta U_1 + \frac{\partial^2 W}{\partial z \partial x} = 0, \quad \varepsilon \Delta U_2 + \frac{\partial^2 W}{\partial z \partial y} = 0, \tag{4}$$

where $\varepsilon = \frac{h}{L} \ll 1, \quad g_1 = \frac{g}{h}$.

The asymptotic solution of (3.1.4) we shall express in the form, which is typical for the method of the geometrical optics:

$$\mathbf{V}(z, x, y) = \sum_{m=0}^{\infty} (i\varepsilon)^m \mathbf{V}_m(z, x, y) e^{\frac{S(x,y)}{i\varepsilon}}, \tag{5}$$

where $\mathbf{V}(z, x, y) = (U_1(z, x, y), U_2(z, x, y), W(z, x, y))$. Functions $S(x, y)$ and $\mathbf{V}_m, m = 0.1,...$ are the subject for determination. Later on we shall confine to the search only for the major member of the expansion (3.1.5) for the vertical component of the velocity $W_0(z, x, y)$. At that from the two last equations of (4) we shall have:

$$U_{10} = -\frac{i \partial S/\partial x}{|\nabla S|^2} \frac{\partial W_0}{\partial z}, \tag{6}$$

$$U_{20} = -\frac{i \partial S/\partial y}{|\nabla S|^2} \frac{\partial W_0}{\partial z}$$

Where: $|\nabla S| = \left(\frac{\partial S}{\partial x}\right)^2 + \left(\frac{\partial S}{\partial y}\right)^2$.

Insert (5) into the first equation of the system (4) and equate the terms of O (1) order

$$\frac{\partial^2 W_0}{\partial z^2} + |\nabla S|^2 \left( \frac{N^2(z, x, y)}{\omega^2} - 1 \right) W_0 = 0, \qquad (7)$$

$$W_0(0, x, y) = W_0(H, x, y) = 0,$$

where $N^2(z, x, y) = \frac{g_1}{\rho_0} \frac{\partial \rho_0}{\partial z}$ - Brunt-Väisälä frequency depending on the horizontal coordinates

For the function $S(x, y)$ we shall have the eikonal equation

$$\left( \frac{\partial S}{\partial x} \right)^2 + \left( \frac{\partial S}{\partial y} \right)^2 = K^2(x, y) \qquad (8)$$

The initial conditions for S eikonal equation for a planar case are set on the link: $L : x_0(\alpha), y_0(\alpha)$

$$S(x, y)|_L = S_0(\alpha).$$

For the solution of the eikonal equation it is necessary to plot the rays, i.e. the characteristics of the equation (.8)

$$\frac{dx}{d\sigma} = \frac{p}{K(x, y)}, \quad \frac{dp}{d\sigma} = \frac{\partial K(x, y)}{\partial x}, \qquad (.9)$$

$$\frac{dy}{d\sigma} = \frac{q}{K(x, y)}, \quad \frac{dq}{d\sigma} = \frac{\partial K(x, y)}{\partial y}$$

where $p = \partial S / \partial x$, $q = \partial S / \partial y$, $d\sigma$ - a ray length element.

The initial conditions $p_0$ and $q_0$ we shall determine from the system of the equations (9)

$$p_0 \frac{\partial x_0}{\partial \alpha} + q_0 \frac{\partial y_0}{\partial \alpha} = \frac{\partial S_0}{\partial \alpha}$$

$$p_0^2 + q_0^2 = K^2(x_0(\alpha), y_0(\alpha))$$

and the initial conditions $x_0(\alpha), y_0(\alpha), p_0(\alpha), q_0(\alpha)$ shall determine the ray $x = x(\sigma, \alpha), y = y(\sigma, \alpha)$. After determination of the rays the eikonal $S$ is being determined by integration over the be

$$S = S_0(\alpha) + \int_0^\sigma K(x(\sigma, \alpha), y(\sigma, \alpha)) d\sigma \qquad (10)$$

Now we shall move to determination of the eigenfunction $W_0(z, x, y)$. We would like to underline, that from (7) it is possible to determine only the vertical dependence of the eigenfunction $W_0(z, x, y)$. In other words, the eigenfunction $W_0$ may be determined with the accuracy up to the multiplication by the arbitrary function $A_0(x, y)$. Now we shall determine $W_0$ in the form of

$$W_0(z, x, y) = A_0(x, y) \hat{W}_0(z, x, y) \qquad (11),$$

where $\hat{W}_0(z, x, y)$ is the solution of the problem (3.1.7) normalized to the following view

$$\int_0^H (N^2(z, x, y) - \omega^2) \hat{W}_0^2(z, x, y) dz = 1. \qquad (12)$$

Let's equate the terms of $O(\varepsilon)$ order after substituting (5) into (4).

$$\omega^2 \left( \frac{\partial^2 W_1}{\partial z^2} + K^2 \left( \frac{N^2(z, x, y)}{\omega^2} - 1 \right) W_1 \right) = (\omega^2 - N^2)(2\nabla W_0 \nabla S + W_0 \Delta S) + \frac{g_1}{\rho_0} (\nabla S \nabla \rho_0) \frac{\partial W_0}{\partial z} - 2(\nabla N^2 \nabla S) W_0$$

,

$$W_1(0, x, y) = W_1(H, x, y).\tag{13}$$

Further we shall take advantage of the condition of orthogonality of the right part of the equation (13) to the function $W_0(z, x, y)$. Multiplying (13) by $W_0$ and integrating by $z$ from 0 up to $H$, we shall receive

$$\int_0^H (N^2(z,x,y) - \omega^2)\nabla(W_0^2 \nabla S)dz - \frac{g_1}{2}\int_0^H (\nabla S \nabla \ln \rho_0)(\frac{\partial W_0^2}{\partial z})dz + 2\int_0^H (\nabla N^2(z,x,y)\nabla S)W_0^2 dz = 0.\tag{14}$$

Transform the second item in (14) using integration by parts and the zero boundary conditions for $W_0$:

$$-\frac{g_1}{2}\int_0^H (\nabla S \nabla \ln \rho_0)(\frac{\partial W_0^2}{\partial z})dz = \frac{\nabla S}{2}\int_0^H \nabla N^2 W_0^2 dz.\tag{15}$$

Transform the first item of (14) considering (12):

$$\int_0^H (N^2 - \omega^2)\nabla(W_0^2 \nabla S)dz = \nabla(A_0^2 \nabla S) - \int_0^H (\nabla S \nabla N^2)W_0^2 dz.\tag{16}$$

To transform the third item of (14), we shall apply the operator of the gradient to the equation (7), considering $Y = \nabla W_0$:

$$\frac{\partial^2 Y(z,x,y)}{\partial z^2} + K^2(x,y)\left(\frac{N^2(z,x,y)}{\omega^2} - 1\right)Y + W_0 \nabla\left[K^2(x,y)\left(\frac{N^2(z,x,y)}{\omega^2} - 1\right)\right] = 0.\tag{17}$$

Multiplying (17) by $W_0$ and integrating by $z$ from 0 up to $H$ and considering (12), we shall receive

$$\int_0^H W_0^2 \nabla N^2(z,x,y)dz = -2A_0^2(x,y)\nabla \ln K(x,y).\tag{18}$$

Finally, we shall rewrite the equation (14) using (15), (16) and (18)

$$\nabla A_0^2 \nabla S + A_0^2 \Delta S - 3\nabla S \nabla \ln K = 0.\tag{.19}$$

The transport equation (19) we shall solve using characteristics of the eikonal equation (9). Using expression for $\Delta S$ along the rays

$$\Delta S = \frac{1}{J}\frac{d}{d\sigma}(JK),$$

where $J(x,y)$ is the geometrical divergence of the rays. Then we shall transform the transport equation (19) to the following law of energy conservation along the rays:

$$\frac{d}{d\sigma}\left(\ln \frac{A_0^2(x,y)J(x,y)}{K^2(x,y)}\right) = 0.\tag{20}$$

The law of the energy conservation (20) can be recorded also in the form suitable for finding $A_0$ function:

$$\frac{A_0^2(x(\sigma,\alpha), y(\sigma,\alpha))}{K^2(x(\sigma,\alpha), y(\sigma,\alpha))}da(x(\sigma,\alpha), y(\sigma,\alpha)) = \frac{A_0^2(x_0(\alpha), y_0(\alpha))}{K^2(x_0(\alpha), y_0(\alpha))}da(x_0(\alpha), y_0(\alpha))\tag{21},$$

where $da(x(\sigma,\alpha), y(\sigma,\alpha)) = J(x(\sigma,\alpha), y(\sigma,\alpha))d\alpha$ is the width of an elementary ray tube. We shall mark, that the current of the wave energy is proportional to $A_0^2 K^{-1} da$, so the (21) shows, that in this case the value equal to the current of the wave energy divided by the modulus of the wave vector is kept safe.

For transition to the study of the problem of the non-harmonic wave packets in the smoothly heterogenous in horizontal direction and the non-stationary stratified medium before selection of ansatzs (Anzatz (german.) - a kind of solution), describing propagation of Airy and Fresnel subinternal waves, let's first consider some leading considerations.

*Airy internal wave.* Inject the slow variables $x^* = \varepsilon x$, $y^* = \varepsilon y$, $t^* = \varepsilon t$ (for $z$ - slowness the variations are not supposed, the character further is neglected), where $\varepsilon = \lambda/L << 1$ is the small parameter characterizing the smoothness of the medium variation in the horizontal direction ($\lambda$ - the characteristic length of the wave, $L$ - the scale of the horizontal heterogeneity). Then the system (2) with the slow variables will look like this:

$$\frac{\partial^4 W}{\partial z^2 \partial t^2} + \varepsilon^2 \frac{\partial^2 W}{\partial t^2} + \frac{g}{\rho_0} \Delta(\varepsilon U_1 \frac{\partial \rho_0}{\partial x} + \varepsilon U_2 \frac{\partial \rho_0}{\partial y} + W \frac{\partial \rho_0}{\partial z}) = 0 \qquad (22)$$

$$\varepsilon \Delta U_1 + \frac{\partial^2 W}{\partial z \partial x} = 0 \quad \varepsilon \Delta U_2 + \frac{\partial^2 W}{\partial z \partial y} = 0 \ .$$

Further, we shall consider the superposition of the cosine waves (with the slow variables $x, y, t$):

$$W = \int \omega \sum_{m=0}^{\infty} (i\varepsilon)^m W_m(\omega, z, x, y) \, e^{\frac{i}{\varepsilon}[\omega t - S_m(\omega, x, y)]} \, d\omega . \qquad (23)$$

Concerning functions $S_m(\omega, x, y)$ it is supposed, that they are odd for $\omega$ and their $\min_\omega \partial S / \partial \omega$ is reached at $\omega = 0$ (for all $x$ and $y$).

Substituting (23) in (22) it will easy to demonstrate, that the function $W_m(\omega, z, x, y)$ at $\omega = 0$ has a pole of m-th order. Therefore the model integrals $R_m(\sigma)$ for the separate units in (23) will be the following expressions:

$$R_m(\sigma) = \frac{1}{2\pi} \int_{-\infty}^{\infty} \left(\frac{i}{\omega}\right)^{m-1} e^{i\left(\frac{\omega^3}{3} - \sigma\omega\right)} d\omega ,$$

where the contour of the integration bends around the dot $\omega = 0$ from above, that ensures the exponential fading of the functions $R_m(\sigma)$ at $\sigma \ll 1$. Functions $R_m(\sigma)$ possess the following property:

$$\frac{d R_m(\sigma)}{d\sigma} = R_{m-1}(\sigma),$$

At that

$$R_0(\sigma) = Ai'(\sigma), \ R_1(\sigma) = Ai(\sigma), \ R_2(\sigma) = \int_{-\infty}^{\sigma} Ai(u) du \text{ etc.}$$

It is obvious, that proceeding from the corresponding properties of Airy integrals, the $R_m(\sigma)$ functions are connected to each other by the following ratios:

$$R_{-1}(\sigma) + \sigma R_1(\sigma) = 0$$
$$R_{-3}(\sigma) + 2 R_0(\sigma) - \sigma^2 R_1(\sigma) = 0 .$$

*Fresnel wave.* As the model integrals $R_m(\sigma)$ describing propagation of Fresnel waves, proceeding from the structure of the solution for the elevation in the horizontally uniform case following expressions are used:

$$R_0(\sigma) = \text{Re} \int_0^\infty \exp\left(-it\sigma - i\frac{t^2}{2}\right) dt \equiv \text{Re} \, \Phi^*(\sigma) \equiv \Phi(\sigma)$$

It quite evident, that function $\Phi^*(\sigma)$ possesses the following property:

$$\frac{d\Phi^*(\sigma)}{d\sigma} = -\int_0^\infty it \exp(-it\sigma - \frac{1}{2}it^2) dt =$$

$$= \int_0^\infty \exp(-it\sigma - \frac{1}{2}it^2) d(-it\sigma - \frac{1}{2}it^2) + i\sigma \int_0^\infty \exp(-it\sigma - \frac{1}{2}it^2) dt = 1 + i\sigma \Phi^*(\sigma)$$

whence, for example, it is possible to obtain:

$$\frac{d^3 \Phi^*(\sigma)}{d\sigma^3} = i\sigma \frac{d^2 \Phi^*(\sigma)}{d\sigma^2} + 2i \frac{d\Phi^*(\sigma)}{d\sigma}$$

or (in terms of functions $R_m(\sigma)$):

$$R_{-1}(\sigma) + i\sigma R_0(\sigma) = 0$$
$$R_{-3}(\sigma) - 2i R_{-1}(\sigma) - i\sigma R_{-2}(\sigma) = 0.$$

Proceeding from the above-stated, and also from the structure of the first item of the uniform asymptotics of Airy and the Fresnel waves in the horizontally uniform medium, the solution of the system (22), for example, it is possible to find in the form of (for the separately taken mode $W_n$ $\mathbf{U}_n$ with the following exclusion of "$n$" index):

$$W = \varepsilon^0 W_0(z,x,y,t) R_0(\sigma) + \varepsilon^a W_1(z,x,y,t) R_1(\sigma) + \varepsilon^{2a} W_2(z,x,y,t) R_2(\sigma) + \mathrm{K} \qquad (24)$$
$$\mathbf{U} = \varepsilon^{1-a} \mathbf{U}_0(z,x,y,t) R_1(\sigma) + \varepsilon \mathbf{U}_1(z,x,y,t) R_2(\sigma) + \varepsilon^{1+a} \mathbf{U}_2(z,x,y,t) R_3(\sigma) + \mathrm{K},$$

where argument $\sigma = \left(\dfrac{1}{a} S(x,y,t)\right)^a \varepsilon^{-a}$ .is considered of the order of unity. Expansion (24) will be in accord with the general approach of the ray optics method and the space-time ray-tracing method.

Let us also to note, that from the similar structure of the solution follows, that in the non-uniform in horizontal direction and the non-stationary medium the solution depends both on the "fast" (the vertical coordinate), and the "slow" (the time and horizontal coordinates) variables. Further we shall look for the solution, as a rule, in the "slow" variables. At that those structural members of the solution, which depends on the "fast" variables, are received in the form of the integrals from some slowly-variable functions along the time-space rays. The given method of the solution allows to present the uniform asymptotes of the fields of the internal gravity waves propagating in the stratified mediums with the slowly-variable parameters, which is true both in the close vicinity and at the far distance from the wave fronts of a separate wave mode.

If it is necessary to describe the behavior of the field only near to the wave front, it is possible to use one of the methods of the ray optics – the method of "traveling wave", and also the low-dispersive approximation in the form of the corresponding local asymptotes and to look for presentation for the argument of the phase functions $\sigma$ in the form of the separate wave mode

$$\sigma = \alpha(t,x,y)(S(t,x,y) - \varepsilon t)\varepsilon^{-a},$$

where the function $S(t,x,y)$ presents the wave front position and is determined from the solution of the eikonal equation

$$\nabla^2 S = c^{-2}(x,y,t)$$

where $c(t,x,y)$ - the maximum group velocity of the corresponding mode, that is the first term of expansion of the dispersing curve in zero. The function $\alpha(t,x,y)$ (the second term of the expansion of the dispersing curve) describes space-time evolution of the pulse width of Airy and Fresnel waves and then will be determined from some energy conservation laws along the characteristics of an eikonal equation, which concrete composition is determined by the physical conditions of the problem.